# Surface versus Impurity Doping Contributions in InAs Nanocrystals Field Effect Transistor Performance

*Durgesh C. Tripathi* [§, ‖], *Lior Asor* [†, ‡, ‖], *Gil Zaharoni* [†, ‡], *Uri Banin*\* [†, ‡]

*and Nir Tessler*\* [§]

§ The Zisapel Nano-Electronics Center, Department of Electrical Engineering, Technion – Israel Institute of Technology, Haifa 32000, Israel

† The Institute of Chemistry, The Hebrew University of Jerusalem, Jerusalem 91904, Israel

‡ The Center for Nanoscience and Nanotechnology, The Hebrew University of Jerusalem, Jerusalem 91904, Israel

ABSTRACT

The electrical functionality of an array of semiconductor nanocrystals depends critically on the free carriers that may arise from impurity or surface doping. Herein, we used InAs nanocrystals thin films as a model system to address the relative contributions of these doping mechanisms by comparative analysis of as-synthesized and Cu-doped nanocrystal based field effect transistor (FET) characteristics. By applying FET simulation methods used in conventional semiconductor FETs, we elucidate surface and impurity-doping contributions to the overall performance of InAs NCs based FETs. As-synthesized InAs nanocrystal based FETs show n-type characteristics assigned to the contribution of surface electrons accumulation layer that can be considered as an actual electron donating doping level with specific doping density and is energetically located just below the conduction band. The Cu-doped InAs NCs FETs show enhanced n-type conduction as expected from the Cu impurities location as an interstitial n-dopant in InAs nanocrystals. The simulated curves reveal the additional contribution from electrons within an impurity sub band close to the conduction band onset of the InAs NCs. The work therefore demonstrates the utility of the bulk FET simulation methodology also to NC based FETs. It provides guidelines for control of doping of nanocrystal arrays separately from surface contributions and impurity doping in colloidal semiconductor NCs towards their future utilization as building blocks in bottom-up prepared optoelectronic devices.

INTRODUCTION

Colloidal semiconductor nanocrystals (NCs) are promising as building blocks for various electronic and optoelectronic applications such as light emitting diodes,[1–7] solar cells,[8–12] field effect transistors (FETs) and more.[13–18] Their functionality in such scenarios benefits from the compatibility with bottom-up device fabrication methods practiced in the general field of "printed electronics" accompanied by the high tunability of their properties afforded by size, shape and composition control.[19] However, impurity doping, commonly practiced in top-down fabricated semiconductor devices to tune the optoelectronic functionality, still presents major challenges in semiconductor nanocrystals based devices.[20–23] In particular–the contribution of impurities to the conduction properties of an array of nanocrystals in a functional device remains an open question. Addressing this will reveal also how many of the introduced impurities are actually electronically active.

In NC based FETs the channel is composed of the colloidal nanocrystals, and addressing to the particle-particle interfaces is critical for achieving high performance devices with high mobility, controlled carrier type and robust current modulation characteristics.[15,18,24,25] An early approach practiced the replacement of the long insulating organic ligands on the as synthesized NC surface with much shorter linker molecules such as diamines and dithiols in order to decrease the distance between the NCs and improve film conductivity. Talapin and Murray reported on high quality FETs with PbSe NCs channels treated by hydrazine which acted both as a linker, and as an electron donor to yield n-type channel devices.[26] Expanding FET functionality to other semiconductor NCs is particularly interesting for III-V semiconductors in relation to their high mobility values and lower toxicity. In the case of InAs NCs based FETs, replacement with short organic ligands such as 1,2 Ethanedithiol and Ethylenediamine led to carrier mobility values on the order of $10^{-5}$ cm$^2$V$^-$

$^1$s$^{-1}$, well below the InAs bulk mobility.[14,27] Significant improvements were reported by replacing the long chain organic ligands with conductive metal chalcogenide complexes (MCCs)[28,29] achieving record mobilities of InAs NC based transistors that reached values as high as 16 cm$^2$V$^{-1}$s$^{-1}$ by using Cu$_7$S$_4^-$ complexes.[30,31] An integrated theoretical-experimental study of such arrays with Sn$_2$S$_6^{4-}$ MCCs as the surface ligands demonstrated that upon film annealing, a nanocomposite is formed with the NCs embedded within the MCCs.[32]

While the control of the NC surface and NC linking strategies in the arrays has yielded FETs exhibiting exceptional mobility values and band-like transport, the utilization of impurity doping in such devices lags significantly behind. The advantage of impurity doping as an additional knob to control the NC device characteristics is obvious. Separate control of NC impurity type and concentration aside from the control of the array mobility by NC surface engineering, will enable a strategy for the bottom-up fabrication of NC based p-n junctions opening the path for a multitude of all-nanocrystal based electrical devices.

Two major routes are relevant for semiconductor nanocrystal doping. Remote, or surface doping is based on either electron donating or accepting molecules or ligands binding to the NCs surface yielding n-type or p-type characteristics.[26,33–35] This type of doping is not fully compatible with NC based devices which themselves require separate tuning of the NC surfaces for controlling device characteristics as discussed above. This calls for the use of impurity doping, which was achieved by either impurity incorporation during synthesis[36,37] or by post-synthesis doping reactions.[22,38–40] The latter approach proved particularly useful for InAs nanocrystals where Cu, Ag and Au impurities were incorporated by a simple room temperature reaction.[22,41–43] Structural analysis using Xray Absorption Spectroscopy methods proved that Cu impurities occupy interstitial lattice sites,[43] in line with the n-type characteristics at the individual nanocrystal level manifested by scanning tunneling spectroscopy, optical spectroscopy and theoretical modelling.[22,44,45]

Addressing the behavior of the Cu –doped InAs NCs in FETs is the focus of this study, aiming to decipher different contributions to the free carriers in the channel. This is a first step in revealing possible contributions to the NC array electrical functionality and to separate surface and ligand related performance from the contribution of impurities. We compared between intrinsic and Cu – doped InAs NCs FETs and analyzed the FETs performance to elucidate the different contributions and factors affecting device characteristics. As-synthesized InAs nanocrystals based FETs show n-type characteristics assigned to the contribution of surface electrons accumulation layer[14,27,46]. Upon Cu doping, the threshold voltage shifts to negative values accompanied by enhanced electron conduction in the gated-regime. Both results indicate contribution of excess electrons due to n-type doping. Interestingly, the mobility of Cu –doped InAs FETs also improved in comparison to intrinsic FETs. In order to further decipher the effects of incorporating doped NCs in the FETs, we utilize formalism adapted from bulk and organic semiconductor FET device analysis to simulate the transfer characteristics of the devices.[47–51] From these simulations, we determined the energetics of the impurity states and the carrier densities. "Intrinsic" (as synthesized, non-intentionally doped NCs) NC devices manifest an electron donating 'doping state' assigned to contribution of surface dangling bonds.[46,52] Devices with Cu-doped InAs NCs manifest additional contribution from an electron donating impurity sub-band that affect both electron density and mobility.

RESULTS & DISCUSSION

A highly monodisperse sample of InAs NCs with a mean diameter of 4.5 nm was used in the fabrication of InAs NCs FETs (for TEM and absorbance spectra see Figure S1 in the supporting information). We fabricated high quality thin film FETs in inert conditions by spin coating a solution of InAs NCs on top of heavily doped Silicon substrates covered with 100 nm thick $SiO_2$ (see experimental section for details). The NCs are capped with bulky insulating organic ligands that were replaced by 1,2 Ethanedithiol (EDT) through a solid state ligand exchange process, to enhance the conductivity by shortening the distance between the NCs. The resulting NCs films were homogeneous and densely packed with an average thickness of 35 nm (see Figure 1b).

We tested a range of thermal annealing treatment conditions prior to electrode evaporation to study its effect on the complete device performance. During the thermal annealing treatment, organic molecules such as solvent residues and unbound ligands desorb from the NCs layer allowing for better coupling between the NCs and enhanced conductivity. Annealing temperatures ranging between $150^0C$ to $220^0C$ under $N_2$ atmosphere were compared and the best performing FET devices were achieved for the highest annealing temperatures. High resolution scanning electron microscope characterization of as-prepared NCs film shows highly dense, amorphous packing of the NCs (Figure 1d). Remarkably, the NCs exhibit thermal stability after being annealed at $220\ ^0C$ and no sign of significant sintering is seen. Energy-dispersive X-ray spectroscopy (EDS) analysis of the film revealed a significant sulfur signature indicating that EDT molecules at least partially remain after annealing as well (Figure S2).

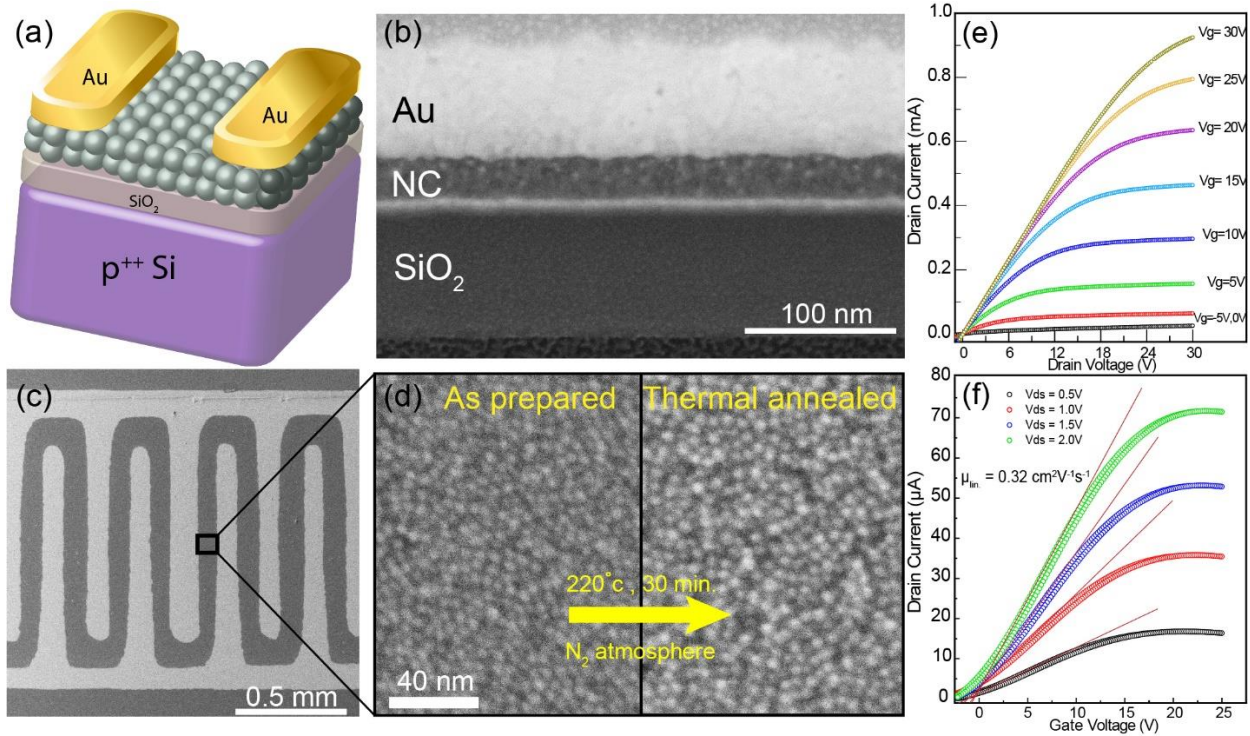

**Figure 1.** (a,b) Illustration and cross section image acquired by high-resolution SEM of the complete device; (c,d) Top view high-resolution SEM image of the NCs layer in between the source and drain electrodes, before and after thermal annealing at 220 $^0$C for 30 minutes under $N_2$ atmosphere; (e,f) Output and transfer characteristics of intrinsic InAs NCs FET cross-linked with EDT.

The intrinsic InAs NCs FETs exhibit clear n-channel current modulation with distinguishable linear and saturation regimes of operation. Figures 1 (e, f) show the electrical characteristics of an intrinsic InAs FET device. The magnitude of drain current at $V_{GS}$=30V approaches 1 mA in the saturation regime and the device exhibits high conductivity with a linear mobility of 0.32 cm$^2$V$^{-1}$s$^{-1}$. This is the highest mobility value reported for FETs of InAs NCs capped with organic ligands and is an order of magnitude lower than the mobility calculated for MCC capped InAs NCs annealed at similar temperature.[31]

With the ability to fabricate high quality InAs NCs films, we next compare between intrinsic InAs and Cu impurity doping contributions in FETs. For that, we synthesized a new batch of InAs NCs and fractions of this batch were doped with Cu impurities. The doping of InAs NCs with Cu atoms, schematically depicted in Figure 2a, follows reported procedures.[22,43] Briefly, Cu salt is solubilized in the presence of surfactants and reducing agents in toluene. Calculated amounts of the Cu solution is mixed with as-synthesized NCs dispersed in toluene at room temperature for 15 minutes. During the reaction, Cu atoms diffuse inside the nanocrystal lattice, where they occupy interstitial sites, as demonstrated in previous works[43] and is illustrated in Figure 2a. We expect that interstitial Cu impurities would donate additional electrons to the NC lattice and increase the free charge carrier density in the NC film.

InAs NCs were doped with varying Cu concentrations. For convenience, we name the samples of InAs NCs doped with a nominal input concentration of 10, 100 and 1000 Cu atoms per NC as low, medium and highly doped samples, respectively. The doped InAs NCs were characterized by TEM, absorption spectroscopy and inductively coupled plasma atomic emission spectroscopy (ICP – AES). ICP – AES measurements were used to determine the actual concentration of Cu atoms in the NCs as summarized in table 1. Absorption spectra of as synthesized and Cu-doped samples is presented together with the shift of the absorption peak relative to the intrinsic sample in Figure 2b. At low Cu doping concentration, the absorption peak is red shifted. This is assigned to the development of Urbach tails, resulting from slight distortions in the NC lattice upon the incorporation of Cu atoms that effectively narrows the band gap.[22] As the amounts of Cu per NC increase, the absorption peak is blue shifted due to states filling next to the conduction band edge by impurity donors, effectively pushing the absorption edge to higher energies. This phenomenon is known as the Moss-Burstein effect and is a strong indication of heavy doping in bulk semiconductors.[53]

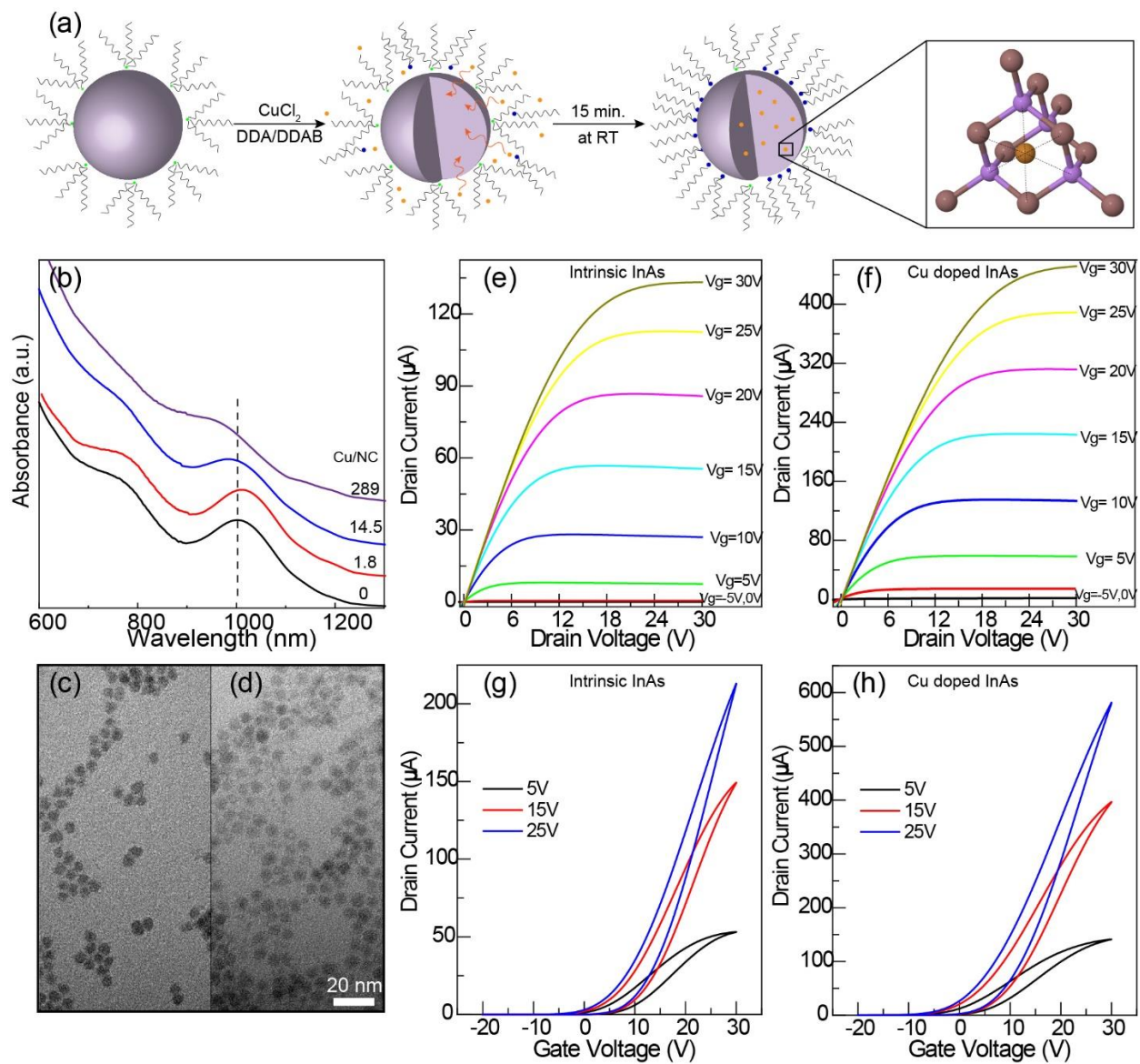

**Figure 2.** (a) Illustration of the doping reaction of InAs NCs with Cu. (b) Absorption spectra of undopedInAs NCs and NCs doped with different amounts of Cu. (c, d) TEM images of 4.5 nm undopedInAs NCs and NCs doped to a level of 289 Cu per NC. (e, f) Measured output characteristics of undoped and 1.8 Cu/NCInAs FETs, respectively. (g, h) Measured transfer characteristics of undoped and 1.8 Cu/NC InAs FETs, respectively.

**Table 1.** Calculated amounts of Cu impurities per NC acquired by ICP-MS

| Doped sample | input Cu/NC concentration | measured Cu/NC concentration |
|---|---|---|
| Intrinsic (undoped) | 0 | 0 |
| Low | 10 | 1.8 |
| Medium | 100 | 14 |
| High | 1000 | 289 |

TEM images of as-synthesized (4.5 nm in diameter) and highly Cu-doped InAs NCs samples (Figure 2(c, d)) indicate that the size and shape of the NCs are well preserved upon doping. Therefore, any effects on optical features of the NCs originate from the incorporation of Cu atoms only.[22] Cu – doped InAs NCs FETs are prepared identically to the undoped InAs FETs as described above. The output characteristics of representative FET devices comprised of a NC film of undoped and low Cu-doped InAs NCs are displayed in Figures 2(e, f), respectively. All devices exhibit n-channel current modulation with distinguishable linear and saturation regimes of operation. The magnitude of the drain current under the same gate bias is found to be higher in Cu-doped devices, in comparison to the as-synthesized InAs NCs. As we'll show below, this trend is due to a variation in the mobility with a contribution from a threshold voltage shift due to enhanced n-type doping.

Devices comprised of InAs NCs doped with a nominal concentration of 100 and 1000 Cu/NC (medium and highly Cu-doped samples) show n-channel current modulation as well. However, the performance of these devices is compromised compared to that of lightly doped samples (See Figures S3, S4, S5). Medium Cu-doped samples also exhibit enhanced conductivity in comparison to as-synthesized NCs FETs, however, in highly Cu–doped devices, the extent of current modulation with gate bias is much weaker with sub-µA magnitude of drain current. The weak current modulation with gate bias is assigned to degenerate doping of those NCs (also evidenced as the Moss Burstein effect in the absorption spectra discussed above). The low current may arise from decreased free carrier mobility that could be assigned to enhanced scattering of the free charge carrier due to high concentration of ionized impurities and lattice disorder that is also manifested in the Urbach tail formation discussed above. From here on, we will focus on InAs NCs FETs made from low Cu-doped NCs samples and compare their results with those of as-synthesized NCs.

Figure 2 (g-h) shows transfer characteristics of the Intrinsic and low Cu – doped InAs FETs, respectively. While all the curves in this study show hysteresis, it is relatively small compared to reports on FETs of similar nanocrystals.[27,54] Moreover, the threshold voltage deduced for the consecutive measurements in positive gate bias remains almost unchanged suggesting the devices recovered from the hysteresis effect (typically a $V_{TH}$ shift) during the hold time (3 sec) between measurements. We used the forward scan of the transfer curves to determine the relevant transistor parameters, analyzing the FETs data in both linear and saturation regimes.

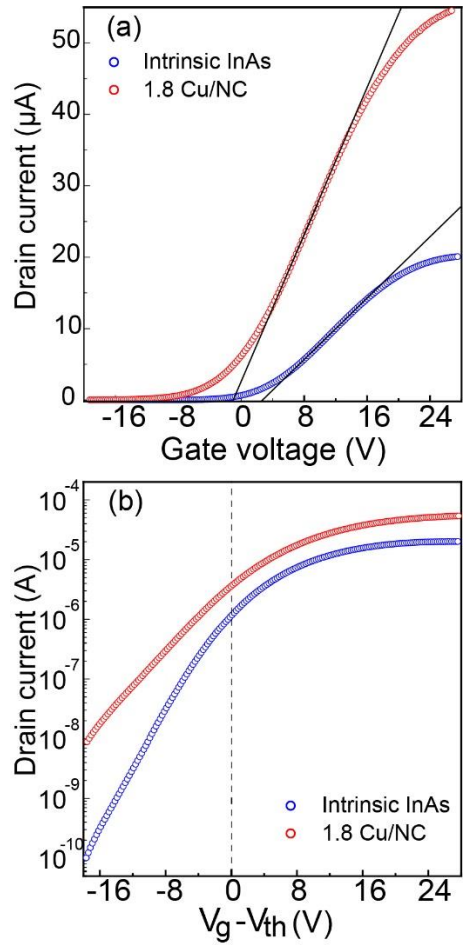

**Figure 3.** (a) Linear regime fittings of intrinsic InAs NC FETs and low Cu – doped InAs FETs at $V_{DS} = 2V$ for mobility and threshold voltage determination; (b) corrected semi-log plot of transfer characteristics.

**Table 2.** Extracted parameters for FETs measured at $V_{ds}=2V$.

| NCs sample | $\mu_{lin}$ (cm$^2$V$^{-1}$s$^{-1}$) | $\mu_{sat}$ (cm$^2$V$^{-1}$s$^{-1}$) | $V_{TH}$ (V) | $I_{on/off}$ | Subthreshold swing (Vdecade$^{-1}$) |
|---|---|---|---|---|---|
| **As-synthesized** | 0.08 | 0.07 | 1.66 | $10^5$ | 4.1 |
| **1.8 Cu/NC** | 0.18 | 0.15 | -2.20 | $10^4$ | 7.1 |

Figure 3(a, b) depicts the transfer characteristics of undoped and low Cu-doped InAs NC based devices measured at a low source-drain voltage ($V_{DS}$=2V). The devices show a well behaved linear regime over a wide voltage range. At higher gate bias, the curves saturate which is the signature of the beginning of hysteresis effect and is beyond the scope of this paper. The linear mobility and threshold voltage extracted in this manner for undoped and lightly doped FETs are summarized in Table 2. Incorporation of impurity dopants in InAs NCs shifts the threshold voltage to negative gate bias values, meaning a smaller shift in the fermi energy is necessary in order to accumulate free carriers in the channel. Threshold voltage tuning due to incorporation of impurity atoms has been reported for Ag-doped PbSe NCs FETs shifting to more positive values as Ag atoms are acting as electron acceptors thus creating holes and pushing the fermi level towards the valence band.[40] In the present case, the more negative threshold voltage indicates that Cu impurities act as electron donors, resulting in the Fermi energy shift towards the conduction band.

Additionally, a two-fold enhancement in the mobility of the Cu-doped films is observed with respect to the undoped devices. As mentioned earlier, carrier mobility may decrease upon impurity doping. A possible explanation for the enhanced mobility is that some impurity ionized charge carriers may passivate intrinsic shallow trap states, facilitating the conductance through the film. This may imply that only a portion of Cu impurities are actually electronically active and contribute free charge carriers to the overall carrier density. We note that the mobility reported in Figure 1 for the as synthesized NCs is somewhat higher than that reported for Figure 2. This reflects differences between various batches and hence, as stated above, we only compared doping levels of the same batch.

In order to correlate our assumptions to the acquired data and better understand the origin of our findings, we performed a two-dimensional simulation study of the transfer characteristics of the FETs made from the undoped and low Cu – doped InAs NCs. We utilized the Sentaurus technology computer aided design (TCAD) simulation platform usually applied to investigate semiconductor devices including organic and perovskite solar cells.[55–57] For the simulations procedures and the material parameters used, see supporting information. The influence of impurity doping on the transfer characteristics is investigated and contribution from different doping mechanisms, including surface electron accumulation layer and impurity doping incorporation, are clarified. Best fits of simulation to measured data are shown in Figure 4.

We started the analysis by simulating the results of the as-synthesized InAs NCs FETs as presented in Figure 4 (a,b) (linear and semi-log scales, respectively). Various doping contributions were included in the simulations. First, NC related effect as InAs NCs are intrinsically n-type and this was assigned to the contribution of a NC surface electron accumulation layer that occurs due to non-passivated surface dangling bonds.[14,46] Second, two device related effects were included: the contribution of fixed surface charge at the insulator-channel interface which modifies the observed threshold voltage, and interfacial hole traps at the same location which affect the subthreshold swing.

All the parameters used in this simulation are listed in Table 3. The effect of the NC shallow surface state doping depends on the density of such states and their energy position relative to the conduction band. Here we set the energy level at 0.15eV below the conduction band and the doping density was fitted. The resulting intrinsic NC 'surface doping' density was found to be $3\times10^{17}$cm$^{-3}$ which is not too high compared to the estimated NC density of $3.3\times10^{18}$ cm$^{-3}$. The effect on the NC surface doping in the simulations is further demonstrated in the inset to Figure 4(a) that shows the simulation results for $V_{DS}$=2V with and without the NC surface doping contribution. Note that the doping indeed contributes to threshold voltage shift, but more importantly for our fitting procedure is that it uniquely affects the curvature of the current around and below $V_{TH}$. Next, we examined the curves on a semi log scale. The inset of Figure 4b shows the fitted curve used for the inset to Figure 4a, but on semi-log scale. It is clear that while the data fits well to simulation in both the gated regime and slightly below $V_{TH}$, it is far from reproducing the experimental curves at the negative gate bias range. Such discrepancy was considered and was shown to be rectified through the introduction of traps at the insulator interface, which is the second device related doping effect in the simulations.[47] We found that introducing a Gaussian distribution of hole traps placed 0.8eV below the conduction band (i.e. below mid-gap) with variance of 0.16 eV and density of $4\times10^{13}$cm$^{-2}$ results in the excellent fit shown in Figure 4b. The effect of these traps can be understood as the following: at negative gate bias the hole traps are gradually filled changing the charge density at the insulator interface such that the effective threshold voltage is shifted to more negative values. Such a shift of $V_{TH}$ in correlation with $V_{GS}$ suppresses the current decline and the measured current persists to very negative gate bias. It should be noted that the interfacial hole traps, introduced to fit the very negative gate bias regime, do not affect the fittings of the rest of the curve. Namely, for the NCs themselves, we were able to deduce the mobility of 0.08 cm$^2$/Vs (Table 2) and the dangling bond induced surface doping to be characterized by energy level of 0.15eV below the conduction band and a density of $3\times10^{17}$cm$^{-3}$.

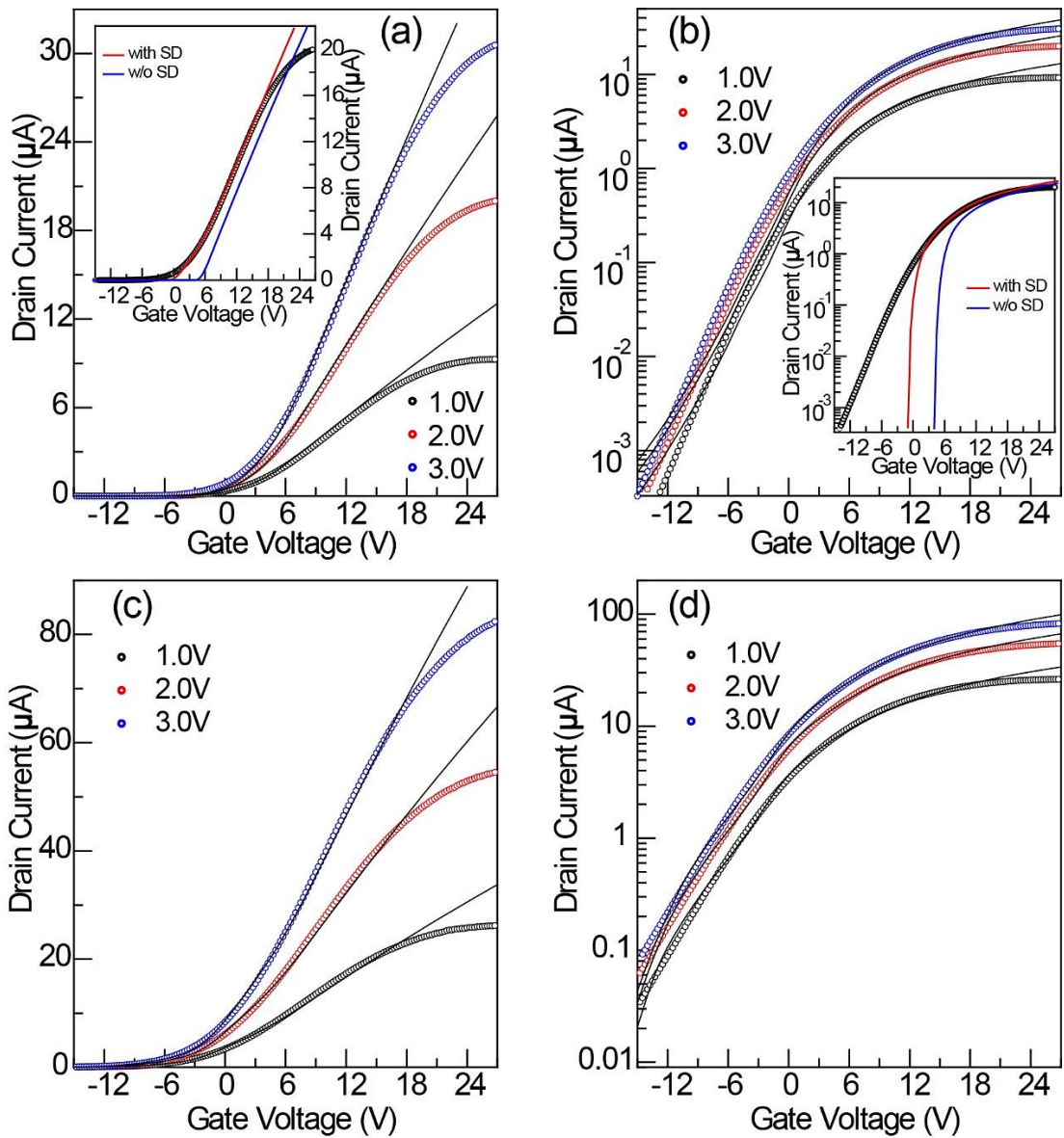

**Figure 4.** Simulated (solid lines) and measured (circles) transfer characteristics of intrinsic InAs NC FETs (a, b) and Cu doped InAs FETs (c,d) for different drain voltages in both linear scale and semi-log scale. Inset in (a) shows fitting with and without NC surface doping (SD) into the gated regime and (b) shows the fitted curve of inset (a) on semi log scale.

**Table 3.** List of simulated parameters used in to fit the transfer curves of undoped and low Cu doped InAs FETs.

|  | Fitting Parameters | Type |  | Undoped | Low Cu doped |
|---|---|---|---|---|---|
| **NCs related effects** | Surface Doping | Single level | Density (cm$^{-3}$) | 3E17 | 3E17 |
|  |  |  | Level relative to Ec (eV) | -0.15 | -0.15 |
|  | Impurity Doping | Gaussian distribution | Density (cm$^{-3}$) | - | 3.7E18 |
|  |  |  | Level relative to Ec (eV) | - | -0.05 |
|  |  |  | Variance (eV) | - | 0.03 |
| **Device related effects** | Interfacial Hole Traps | Gaussian distribution | Density (cm$^{-2}$) | 4E13 | 7E12 |
|  |  |  | Level relative to Ec (eV) | -0.8 | -0.8 |
|  |  |  | Variance (eV) | 0.16 | 0.16 |
|  | Insulator Fixed Charge | Negative charge | Density (cm$^{-2}$) | 5E11 | 5E11 |

Next, we discuss the simulation of the data for the low Cu-doped (1.8 Cu/NC) InAs FETs transfer curves, relying on the results obtained for the undoped case. Starting with the linear scale curves (Figure 4c) we retain the NC intrinsic surface doping and the fixed charge at the insulator interface and allow for an additional impurity doping level that would represent the NC effect of the dopant Cu atoms. Namely, we assume that Cu doping does not significantly change the density of dangling bonds as it is incorporated as an interstitial dopant within the NC lattice. We chose the energy position of these doping levels based on STM data on such Cu doped NCs,[22] and placed it 0.05eV below the conduction band and allowed it to be somewhat broader than just a single level (i.e. Gaussian with 0.03eV variance). The excellent fits are shown in Figure 4c and by fitting the curves, we found the dopant density to be $3.7 \times 10^{18} cm^{-3}$. Looking more carefully in the semi-log curves (Figure 4d) we again had to include hole traps at the insulator interface so as to reproduce the negative gate bias range. For best fit we used the hole traps deduced for the undoped case and their density is reduced (see Table 3).

To shed more light on the transistor's operation, we used the above-mentioned parameters for both undoped and low Cu - doped samples to calculate the conduction band energy, electron quasi-Fermi energy and carrier density across the NC layer in the middle of the channel at three different bias values: in the deep subthreshold regime, at threshold, and in the deep gated regime (see Figure 5). When the gate bias is at the subthreshold regime, the free charge carriers at the NC layer close to the oxide–semiconductor interface are depleted. Once the gate bias passes above the threshold voltage and enters the gated regime, free electrons accumulate at the semiconductor–oxide interface. The charge density far from the oxide can be taken as a measure for the equilibrium charge density of the semiconductor layer. Using the results for $V_G=V_{TH}$, the free carrier density at the surface of the NC layer is estimated to be $1.4\times10^{17}$ cm$^{-3}$ and $3.1\times10^{17}$ cm$^{-3}$ for as-synthesized and low Cu doped FETs, respectively. From this, one can infer that the impurity doping generates additional $1.7\times10^{17}$ cm$^{-3}$ charge carriers in the low Cu doped FETs sample.

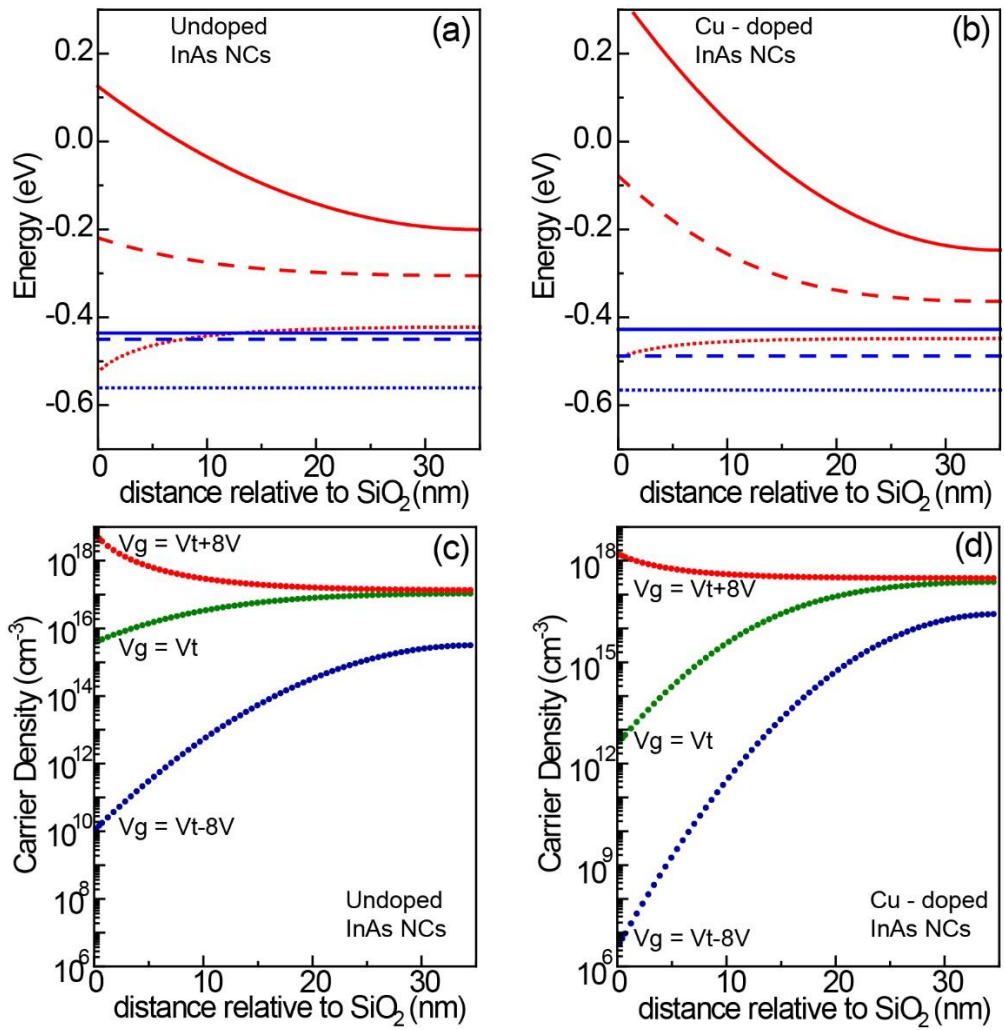

**Figure 5.** (a, b) Conduction band energy (Red curves) and electron quasi-Fermi energy (Blue lines) variation with depth of the semiconductor NC layer of undoped and Cu – doped InAs FETs, respectively, at three different effective gate voltages, representing different operating regimes of the FET: deep subthreshold ($V_G=V_{TH}-8V$, solid lines), threshold ($V_{TH}$, dashed lines), and gated regime ($V_{TH}+8V$, dash-dotted lines). (c, d) Carrier density variation versus depth into the semiconductor NC layer of undoped and Cu – doped InAs FETs, respectively, at three different effective gate voltages.

Using these detailed simulations, we found that the FET characteristics exhibit intrinsic effects associated with the nanocrystals as well as extrinsic effects associated with the device insulator interface. This simulation is based on general principles and can be applied to wide range of materials. With the help of the simulations, we were able to "clean" the curves and delineate intrinsic and impurity doping contributions to the NCs film conductivity and demonstrate the origin of a significant improvement in charge carrier mobility in low Cu – doped InAs NCs FETs.

Prior to further discussing the implications of the specific FET device results simulations, we consider the merit of the simulation methodology. Bulk doping has two effects on transistor characteristics. It shifts the threshold voltage and softens the onset of the current above the threshold voltage. We found that this softening is a unique signature to the doping contribution and hence this was the main feature used to extract the NC doping density. The actual position of $V_{TH}$ was then adjusted using fixed surface charge. Also, the method of finding the gate voltage at which the semiconductor is fully depleted may be hindered by the presence of traps at the insulator interface. Only by fitting the full range using a TCAD tool one can fully account for the role of such traps. Regarding parameter extraction, we note that the saturation regime of the transfer characteristics is not reliable in the presence of doping. In this range, as the gate bias is increased, the region close to the insulator interface is gradually changing from depletion to accumulation (see Figure 5c, d). Namely, in this saturation regime, the channel does not solely govern the current and one should not use it to deduce either mobility or threshold voltage.

Finally, recapping and focusing the attention to the NCs - we found that undoped NCs contain $3\times10^{17} cm^{-3}$ surface related electron donating states. We also found that introducing about 2 Cu atoms per NC results in a net dopant density of $3.7\times10^{18} cm^{-3}$. Using the estimated Cu atoms density ($7\times10^{18}$ cm$^{-3}$) we find that only ~50% of the Cu atoms can perform as dopants, Namely, it is not enough that Cu will penetrate the NC but it needs to be in a certain configuration to act as dopant. The above is not to be confused with the fact that not all dopants are ionized. Under flat-band conditions we found that the total density of free charge carriers is $1.4\times10^{17}$cm$^{-3}$ and $3.1\times10^{17}$cm$^{-3}$ for the undoped and low Cu – doped NCs, respectively. Namely, only about $1.7\times10^{17}$cm$^{-3}$ Cu dopants were ionised at the room temperature conditions of these measurements. It is reasonable to assume that some of the electronically inactive impurities may passivate shallow electron trap states, leading to the observed enhancement in the mobility of the NCs film.

CONCLUSION

In conclusion, using a combination of experiments and simulations of the InAs NC FETs characteristics, we have investigated the NC surface versus impurity doping contributions. As-synthesized NCs exhibit n-type characteristics assigned to contribution of surface electron accumulation in traps. Upon doping the NCs with Cu the mobility is enhanced (by ~150%) and the threshold voltage shifts to negative values. These effects are well represented by the addition of n-type impurity doping sub-band contributing to the overall charge carrier density. The improved mobility is assigned to NC surface trap passivation. While this indicates the potential for possible separate control of doping type and density by the introduction of impurity dopants in NCs, there is still close relation between surface effects and impurity effects in their behavior. Further work may enable further separation of the effects of surface versus impurity contributions allowing tailoring of the electrical characteristics of NC based devices.

METHODS

**Synthesis of colloidal InAs NCs.** Colloidal InAs nanocrystals were synthesized following a well-established wet-chemical synthesis.[58] Precursor solutions containing mixtures of (TMS)$_3$As (tris(trimethylsilyl)arsine) and InCl$_3$ (indium(III) chloride) were prepared and kept under inert conditions throughout. At the schlenk line, a solution of distilled TOP (tri-n-octylphosphine) was evacuated for 30 min and then heated to 300 °C. The nucleation solution (molar ratio 2:1 In:As) was rapidly injected, and the heating mantle was removed, lowering the solution temperature to 260 °C. The growth solution (1.2:1 In:As) was gradually introduced to the solution, allowing for particle growth until the desired size was reached. Narrow size distributions were achieved by size selective precipitation performed in a glove box by adding methanol to the NC dispersion and filtering the solution through a 0.2 μm polyamide membrane filter (Whatman). The precipitant was subsequently dissolved in anhydrous-toluene (Sigma) and kept under constant inert conditions throughout.

**Doping Presynthesized InAs Nanocrystals**. Doping presynthesized InAs NCs was achieved following a room-temperature solution-phase reaction previously reported by us.[22] In short, an impurity solution was prepared by dissolving the metal salt, that is, copper chloride (CuCl$_2$), didodecyldimethylammonium bromide (DDAB), and dodecylamine (DDA) in anhydrous-toluene (Sigma). The solution was agitated using an ultrasonic bath for ~15 min to ensure homogeneous conditions. Calculated amounts (v/v) of the impurity solution were added to InAs NCs suspensions in toluene. The doping reaction was performed under inert condition inside the glove box while stirring. The reaction was terminated by adding methanol and separating the precipitated doped NCs from solution with centrifugation (5 min, 3000 rpm). The supernatant containing the unreacted impurities was discarded and the doped NC solution was re-dissolved in hexane. We note that the Cu/NC ratios reported in the manuscript correspond to the ratio of Cu impurities per NCs for the doping reaction in solution. The actual Cu concentration per NC is derived from this and determined as detailed next.

**ICP-AES Measurements**. Inductively coupled plasma − atomic emission spectroscopy (ICP-AES) measurements were carried out using a Perkin-Elmer Optima 3000 to determine dopant concentrations within the NCs. The emission signals were collected at 230.6 and 325.6 nm for In ions, 189.0 and 193.7 nm for As ions, and 324.8 and 328.1 nm for Cu. Samples were prepared by dissolving the doped NCs in HNO3 and diluting the solution with triply distilled water.

**Field Effect Transistors fabrication and Electrical Measurements**. Heavily doped p-type Si substrates covered with 100 nm thick $SiO_2$ were cleaned as follows: The substrate were dipped into Acetone and put to sonication for 5 min. following that, the substrates were dipped in methanol and again sonicated for 5 min. for the last run, the substrates were dipped in isopropanol and sonicated for 5 min. After this cleaning procedure, the substrates were blown dry with $N_2$ and put inside a plasma cleaner for 30 min. in the glove box, 0.005 M (3-mercaptopropyl) trimetoxysilane (3-MPTMS) in methanol was prepared, and the substrates were kept in this solution overnight. Thin films of InAs NCs on top of $Si/SiO_2$ substrates were prepared as follows: 20 mg/ml solution of InAs NCs in hexane were prepared as described above and filtered twice to remove large aggregates and impurities. 30 μL of clean InAs NCs solution was dropped on $Si/SiO_2$ substrate and span at 2000 RPM for 30 seconds. Solid state ligand exchange with 1,2 Ethanedithiol (EDT) was performed by covering the film with 0.005 M solution of EDT in Acetonitrile for 30 second, and spinning at 2000 rpm for 15 seconds to dry the film. The film was washed by acetonitrile twice with spinning the substrate at 2000 rpm for 30 seconds. The entire cycle was repeated twice more to prepare a 30-35 nm thick film of InAs NCs. As prepared films of InAs NCs were put inside an oven inside the glovebox and heated to 220c for 30 min under $N_2$ environment, to remove organic residues. 120 nm thick Au Source/Drain electrodes (interdigitated configuration; W = 20 mm, L = 100 μm) were thermally evaporated using a shadow mask by the thermal evaporation. Immediately after finishing the device fabrication, electrical measurements were performed using the semiconductor parameter analyzer (Agilent-4200). All device fabrication steps and electrical measurements were performed in the $N_2$-filled gloveboxes.

## ASSOCIATED CONTENT

**Supporting information**

Additional experimental details, methodology of FET simulation, TEM images and electrical characterization of FETs are found in the supporting information file.

This material is available free of charge via the Internet at http://pubs.acs.org.

## AUTHOR INFORMATION

**Corresponding Authors**

*Emails: nir@ef.technion.ac.il; uri.banin@mail.huji.ac.il

**Author Contributions**

‖L.A. and D.C.T. contributed equally to this work.

All authors have given approval to the final version of the manuscript.

**Notes**

The authors declare no competing financial interest.

ACKNOWLEDGEMENTS

This research was supported by the Israeli Innovation Authority under KAMIN program (No. 57711 & 61266) and by the NSF-BSF International Collaboration in Chemistry program (NSF Grant No. CHE-1413937 and BSF Grant No. 2013/610). Additional partial support was provided by the ISF-NSFC joint research program (grant No.2495/17) (U.B.) and by Israel Science Foundation (grant no. 488/16) (N.T.) . U.B. thanks the Alfred & Erica Larisch memorial chair.

REFERENCES

(1)     Tessler, N.; Medvedev, V.; Kazes, M.; Kan, S. H.; Banin, U. Efficient Near-Infrared Polymer Nanocrystal Light-Emitting Diodes. *Science.* **2002**, *295*, 1506–1508.

(2)     Ahn, J. H.; Bertoni, C.; Dunn, S.; Wang, C.; Talapin, D. V.; Gaponik, N.; Eychmüller, A.; Hua, Y.; Bryce, M. R.; Petty, M. C. White Organic Light-Emitting Devices Incorporating Nanoparticles of II-VI Semiconductors. *Nanotechnology* **2007**, *18*, 335202.

(3)     Caruge, J. M.; Halpert, J. E.; Wood, V.; Bulovíc, V.; Bawendi, M. G. Colloidal Quantum-Dot Light-Emitting Diodes with Metal-Oxide Charge Transport Layers. *Nat. Photonics* **2008**, *2*, 247–250.

(4)     Anikeeva, P. O.; Halpert, J. E.; Bawendi, M. G.; Bulović, V. Quantum Dot Light-Emitting Devices with Electroluminescence Tunable over the Entire Visible Spectrum. *Nano Lett.* **2009**, *9*, 2532–2536.

(5)     Talapin, D. V.; Steckel, J. Quantum Dot Light-Emitting Devices. *MRS Bull.* **2013**, *38*, 685–691.

(6)     Gong, X.; Yang, Z.; Walters, G.; Comin, R.; Ning, Z.; Beauregard, E.; Adinolfi, V.; Voznyy, O.; Sargent, E. H. Highly Efficient Quantum Dot Near-Infrared Light-Emitting Diodes. *Nat. Photonics* **2016**, *10*, 253–257.

(7)     Pradhan, S.; Di Stasio, F.; Bi, Y.; Gupta, S.; Christodoulou, S.; Stavrinadis, A.; Konstantatos, G. High-

Efficiency Colloidal Quantum Dot Infrared Light-Emitting Diodes via Engineering at the Supra-Nanocrystalline Level. *Nat. Nanotechnol.* **2019**, *14*, 72–79.

(8) Tang, J.; Liu, H.; Zhitomirsky, D.; Hoogland, S.; Wang, X.; Furukawa, M.; Levina, L.; Sargent, E. H. Quantum Junction Solar Cells. *Nano Lett.* **2012**, *12*, 4889–4894.

(9) Tvrdy, K.; Kamat, P. V. Quantum Dot Solar Cells. *Compr. Nanosci. Technol.* **2011**, *4*, 257–275.

(10) Ning, Z.; Zhitomirsky, D.; Adinolfi, V.; Sutherland, B.; Xu, J.; Voznyy, O.; Maraghechi, P.; Lan, X.; Hoogland, S.; Ren, Y.; et al. Graded Doping for Enhanced Colloidal Quantum Dot Photovoltaics. *Adv. Mater.* **2013**, *25*, 1719–1723.

(11) Leschkies, K. S.; Beatty, T. J.; Kang, M. S.; Norris, D. J.; Aydil, E. S. Solar Cells Based on Junctions between Colloidal Pbse Nanocrystals and Thin ZnO Films. *ACS Nano* **2009**, *3*, 3638–3648.

(12) Chuang, C. H. M.; Brown, P. R.; Bulović, V.; Bawendi, M. G. Improved Performance and Stability in Quantum Dot Solar Cells through Band Alignment Engineering. *Nat. Mater.* **2014**, *13*, 796–801.

(13) Hetsch, F.; Zhao, N.; Kershaw, S. V.; Rogach, A. L. Quantum Dot Field Effect Transistors. *Mater. Today* **2013**, *16*, 312–325.

(14) Geyer, S. M.; Allen, P. M.; Chang, L. Y.; Wong, C. R.; Osedach, T. P.; Zhao, N.; Bulovic, V.; Bawendi, M. G. Control of the Carrier Type in InAs Nanocrystal Films by Predeposition Incorporation of Cd. *ACS Nano* **2010**, *4*, 7373–7378.

(15) Choi, J. H.; Fafarman, A. T.; Oh, S. J.; Ko, D. K.; Kim, D. K.; Diroll, B. T.; Muramoto, S.; Gillen, J. G.; Murray, C. B.; Kagan, C. R. Bandlike Transport in Strongly Coupled and Doped Quantum Dot Solids: A Route to High-Performance Thin-Film Electronics. *Nano Lett.* **2012**, *12*, 2631–2638.

(16) Choi, J. H.; Wang, H.; Oh, S. J.; Paik, T.; Jo, P. S.; Sung, J.; Ye, X.; Zhao, T.; Murray, C. B.; Kagan, C. R.; et al. Exploiting the Colloidal Nanocrystal Library to Construct Electronic Devices. *Science.* **2016**, *352*, 205–208.

(17) Kagan, C. R. Flexible Colloidal Nanocrystal Electronics. *Chem. Soc. Rev.* **2019**, *48*, 1626–1641.

(18)  Talapin, D. V.; Lee, J. S.; Kovalenko, M. V.; Shevchenko, E. V. Prospects of Colloidal Nanocrystals for Electronic and Optoelectronic Applications. *Chem. Rev.* **2010**, *101*, 389-458.

(19)  Kovalenko, M. V.; Manna, L.; Cabot, A.; Hens, Z.; Talapin, D. V.; Kagan, C. R.; Klimov, V. I.; Rogach, A. L.; Reiss, P.; Milliron, D. J.; et al. Prospects of Nanoscience with Nanocrystals. *ACS Nano* **2015**, *9*, 1012–1057.

(20)  Buonsanti, R.; Milliron, D. J. Chemistry of Doped Colloidal Nanocrystals. *Chem. Mater.* **2013**, *25*, 1305–1317.

(21)  Norris, D. J.; Efros, A. L.; Erwin, S. C. Doped Nanocrystals. *Science.* **2008**, *319*, 1776–1779.

(22)  Mocatta, D.; Cohen, G.; Schattner, J.; Millo, O.; Rabani, E.; Banin, U. Heavily Doped Semiconductor Nanocrystal Quantum Dots. *Science.* **2011**, *332*, 77–81.

(23)  Stavrinadis, A.; Konstantatos, G. Strategies for the Controlled Electronic Doping of Colloidal Quantum Dot Solids. *ChemPhysChem.* **2016**, *17*, 632–644.

(24)  Kagan, C. R.; Murray, C. B. Charge Transport in Strongly Coupled Quantum Dot Solids. *Nat. Nanotechnol.* **2015**, *10*, 1013–1026.

(25)  Oh, S. J.; Straus, D. B.; Zhao, T.; Choi, J. H.; Lee, S. W.; Gaulding, E. A.; Murray, C. B.; Kagan, C. R. Engineering the Surface Chemistry of Lead Chalcogenide Nanocrystal Solids to Enhance Carrier Mobility and Lifetime in Optoelectronic Devices. *Chem. Commun.* **2017**, *53*, 728–731.

(26)  Talapin, D. V.; Murray, C. B. PbSe Nanocrystal Solids for N- and p-Channel Thin Film Field-Effect Transistors. *Science.* **2005**, *310*, 86–89.

(27)  Soreni-Harari, M.; Mocatta, D.; Zimin, M.; Gannot, Y.; Banin, U.; Tessler, N. Interface Modifications of InAs Quantum-Dots Solids and Their Effects on FET Performance. *Adv. Funct. Mater.* **2010**, *20*, 1005–1010.

(28)  Kovalenko, M. V.; Scheele, M.; Talapin, D. V. Colloidal Nanocrystals with Molecular Metal Chalcogenide Surface Ligands. *Science.* **2009**, *324*, 1417–1420.

(29) Lee, J.-S.; Kovalenko, M. V.; Huang, J.; Chung, D. S.; Talapin, D. V. Band-like Transport, High Electron Mobility and High Photoconductivity in All-Inorganic Nanocrystal Arrays. *Nat. Nanotechnol.* **2011**, *6*, 348–352.

(30) Jang, J.; Liu, W.; Son, J. S.; Talapin, D. V. Temperature-Dependent Hall and Field-Effect Mobility in Strongly Coupled All-Inorganic Nanocrystal Arrays. *Nano Lett.* **2014**, *14*, 653–662.

(31) Liu, W.; Lee, J. S.; Talapin, D. V. III-V Nanocrystals Capped with Molecular Metal Chalcogenide Ligands: High Electron Mobility and Ambipolar Photoresponse. *J. Am. Chem. Soc.* **2013**, *135*, 1349–1357.

(32) Scalise, E.; Srivastava, V.; Janke, E.; Talapin, D.; Galli, G.; Wippermann, S. Surface Chemistry and Buried Interfaces in All-Inorganic Nanocrystalline Solids. *Nat. Nanotechnol.* **2018**, *13*, 841–848.

(33) Shim, M.; Guyot-Sionnest, P. N-Type Colloidal Semiconductor Nanocrystals. *Nature.* **2000**, *407*, 981–983.

(34) Koh, W. K.; Koposov, A. Y.; Stewart, J. T.; Pal, B. N.; Robel, I.; Pietryga, J. M.; Klimov, V. I. Heavily Doped N-Type PbSe and PbS Nanocrystals Using Ground-State Charge Transfer from Cobaltocene. *Sci. Rep.* **2013**, *3*, 2004.

(35) Kirmani, A. R.; Kiani, A.; Said, M. M.; Voznyy, O.; Wehbe, N.; Walters, G.; Barlow, S.; Sargent, E. H.; Marder, S. R.; Amassian, A. Remote Molecular Doping of Colloidal Quantum Dot Photovoltaics. *ACS Energy Lett.* **2016**, *1*, 922–930.

(36) Sahu, A.; Kang, M. S.; Kompch, A.; Notthoff, C.; Wills, A. W.; Deng, D.; Winterer, M.; Frisbie, C. D.; Norris, D. J. Electronic Impurity Doping in CdSe Nanocrystals. *Nano Lett.* **2012**, *12*, 2587–2594.

(37) Capitani, C.; Pinchetti, V.; Gariano, G.; Santiago-González, B.; Santambrogio, C.; Campione, M.; Prato, M.; Brescia, R.; Camellini, A.; Bellato, F.; et al. Quantized Electronic Doping towards Atomically Controlled "Charge-Engineered" Semiconductor Nanocrystals. *Nano Lett.* **2019**, *19*, 1307–1317.

(38) Kroupa, D. M.; Hughes, B. K.; Miller, E. M.; Moore, D. T.; Anderson, N. C.; Chernomordik, B. D.; Nozik, A. J.; Beard, M. C. Synthesis and Spectroscopy of Silver-Doped PbSe Quantum Dots. *J. Am. Chem. Soc.* **2017**, *139*, 10382–10394.

(39) Lu, H.; Carroll, G. M.; Chen, X.; Amarasinghe, D. K.; Neale, N. R.; Miller, E. M.; Sercel, P. C.; Rabuffetti, F. A.; Efros, A. L.; Beard, M. C. N-Type PbSe Quantum Dots via Post-Synthetic Indium Doping. *J. Am. Chem. Soc.* **2018**, *140*, 13753–13763.

(40) Kang, M. S.; Sahu, A.; Frisbie, C. D.; Norris, D. J. Influence of Silver Doping on Electron Transport in Thin Films of PbSe Nanocrystals. *Adv. Mater.* **2013**, *25*, 725–731.

(41) Amit, Y.; Li, Y.; Frenkel, A. I.; Banin, U. From Impurity Doping to Metallic Growth in Diffusion Doping: Properties and Structure of Silver-Doped InAs Nanocrystals. *ACS Nano* **2015**, *9*, 10790–10800.

(42) Liu, J.; Amit, Y.; Li, Y.; Plonka, A. M.; Ghose, S.; Zhang, L.; Stach, E. A.; Banin, U.; Frenkel, A. I. Reversed Nanoscale Kirkendall Effect in Au-InAs Hybrid Nanoparticles. *Chem. Mater.* **2016**, *28*, 8032–8043.

(43) Amit, Y.; Eshet, H.; Faust, A.; Patllola, A.; Rabani, E.; Banin, U.; Frenkel, A. I. Unraveling the Impurity Location and Binding in Heavily Doped Semiconductor Nanocrystals: The Case of Cu in InAs Nanocrystals. *J. Phys. Chem. C.* **2013**, *117*, 13688–13696.

(44) Yang, C.; Faust, A.; Amit, Y.; Gdor, I.; Banin, U.; Ruhman, S. Impurity Sub-Band in Heavily Cu-Doped InAs Nanocrystal Quantum Dots Detected by Ultrafast Transient Absorption. *J. Phys. Chem. A* **2016**, *120*, 3088–3097.

(45) Faust, A.; Amit, Y.; Banin, U. Phonon-Plasmon Coupling and Active Cu Dopants in Indium Arsenide Nanocrystals Studied by Resonance Raman Spectroscopy. *J. Phys. Chem. Lett.* **2017**, *8*, 2519–2525.

(46) Noguchi, M.; Hirakawa, K.; Ikoma, T. Intrinsic Electron Accumulation Layers on Reconstructed Clean InAs(100) Surfaces. *Phys. Rev. Lett.* **1991**, *66*, 2243-2246.

(47) Scheinert, S.; Paasch, G.; Schrödner, M.; Roth, H. K.; Sensfuß, S.; Doll, T. Subthreshold Characteristics of Field Effect Transistors Based on Poly(3-Dodecylthiophene) and an Organic Insulator. *J. Appl. Phys.* **2002**, *92*, 330–337.

(48) Scheinert, S.; Paasch, G. Fabrication and Analysis of Polymer Field-Effect Transistors. *Phys. Status Solidi Appl. Res.* **2004**, *201*, 1263–1301.

(49) Scheinert, S.; Pernstich, K. P.; Batlogg, B.; Paasch, G. Determination of Trap Distributions from Current Characteristics of Pentacene Field-Effect Transistors with Surface Modified Gate Oxide. *J. Appl. Phys.* **2007**, *102*, 104503.

(50) Possanner, S. K.; Zojer, K.; Packer, P.; Zojer, E.; Schürrer, F. Threshold Voltage Shifts in Organic Thin-Film Transistors Due to Self-Assembled Monolayers at the Dielectric Surface. *Adv. Funct. Mater.* **2009**, *19*, 958–967.

(51) Völkel, A. R.; Street, R. A.; Knipp, D. Carrier Transport and Density of State Distributions in Pentacene Transistors. *Phys. Rev. B - Condens. Matter Mater. Phys.* **2002**, *66*, 1953361–1953368.

(52) Hang, Q.; Wang, F.; Carpenter, P. D.; Zemlyanov, D.; Zakharov, D.; Stach, E. A.; Buhro, W. E.; Janes, D. B. Role of Molecular Surface Passivation in Electrical Transport Properties of InAs Nanowires. *Nano Lett.* **2008**, *8*, 49–55.

(53) Abram, R. A.; Rees, G. J.; Wilson, B. L. H. Heavily Doped Semiconductors and Devices. *Adv. Phys.* **1978**, *27*, 799–892.

(54) Song, J. H.; Choi, H.; Pham, H. T.; Jeong, S. Energy Level Tuned Indium Arsenide Colloidal Quantum Dot Films for Efficient Photovoltaics. *Nat. Commun.* **2018**, *9*, 1–9.

(55) Alnuaimi, A.; Almansouri, I.; Nayfeh, A. Performance of Planar Heterojunction Perovskite Solar Cells under Light Concentration. *AIP Adv.* **2016**, *6,* 115012.

(56) Würfel, U.; Neher, D.; Spies, A.; Albrecht, S. Impact of Charge Transport on Current-Voltage Characteristics and Power-Conversion Efficiency of Organic Solar Cells. *Nat. Commun.* **2015**, *6*, 1–9.

(57) www.synopsys.com. Sentaurus Device User Guide J-2014.09. **2014**, No. September, 1488.

(58) Cao, Y. W.; Banin, U. Growth and Properties of Semiconductor Core/Shell Nanocrystals with InAs Cores. *J. Am. Chem. Soc.* **2000**, *122*, 9692–9702.

**Table of Content figure:**

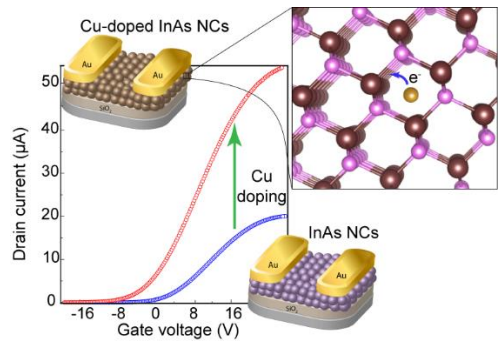

# Supporting Information

# Surface versus Impurity Doping Contributions in InAs Nanocrystal Field Effect Transistor Performance

*Durgesh C. Tripathi* [§, ‖], *Lior Asor* [†, ‡, ‖], *Gil Zaharoni* [†, ‡], *Uri Banin*\* [†, ‡]

*and Nir Tessler*\* [§]

§ The Zisapel Nano-Electronics Center, Department of Electrical Engineering, Technion – Israel Institute of Technology, Haifa 32000, Israel

† The Institute of Chemistry, the Hebrew University of Jerusalem, Jerusalem 91904, Israel

‡ The Center for Nanoscience and Nanotechnology, the Hebrew University of Jerusalem, Jerusalem 91904, Israel

# Experimental section

**Materials.** Toluene (anhydrous, 99.8%), hexane (anhydrous, 99.8%), Methanol (anhydrous, 99.8%), acetonitrile (anhydrous, 99.8%), (3-mercaptopropyl)trimethoxysilane (95%), 1,2-Ethanedithiol (98%), Indium(III) chloride (99.999%), Copper(II) chloride (99.995%), Dodecylamine (98%), didodecyldimethylammonium bromide (98%) were purchased from Sigma-Aldrich (Merck) and used as received. Tri-n-octylphosphine (90%) was purchased from Sigma-Aldrich (Merck) and was distilled prior to use. Tris(trimthylsylil)arsine ((TMS)$_3$As) was synthesized in our lab.

## Simulation Method and Standard Material Parameters

A diagram of the simulated NC field effect transistor is illustrated schematically in Figure S6. The thickness of the oxide layer is taken as 100 nm and the NC semiconductor layer thickness is 35 nm. The channel length (i.e. the distance between the source and the drain electrodes) is 5 µm and the channel width 1000 µm. Thus, the W/L ratio of the simulated structure is 200, similar to the experimental W/L ratio. In the field effect transistor, the application of strong positive gate bias induces a two-dimensional electron sheet in the semiconductor layer just above the oxide layer. The concentration of charge carriers at the semiconductor - oxide interface and the larger W/L ratio, justify performing a 2D simulation study of the charge transport in the FET structure.

The simulation study is performed by solving the standard drift-diffusion, Poisson, and continuity equations for the $I_{DS}$-$V_{GS}$ transfer characteristics of the fabricated devices on the Senatrus TCAD platform. It is reported that the undoped InAs NC have excess n-type of charge carriers due to dangling bonds located on the NC surfaces. In our simulation, the surface doping level is placed 0.15eV below the conduction band. The other parameters taken into this simulation study are summarized in Table S1. For the mobility, experimentally determined values are used which is given in Table 2 of the main paper.

# Figures and tables

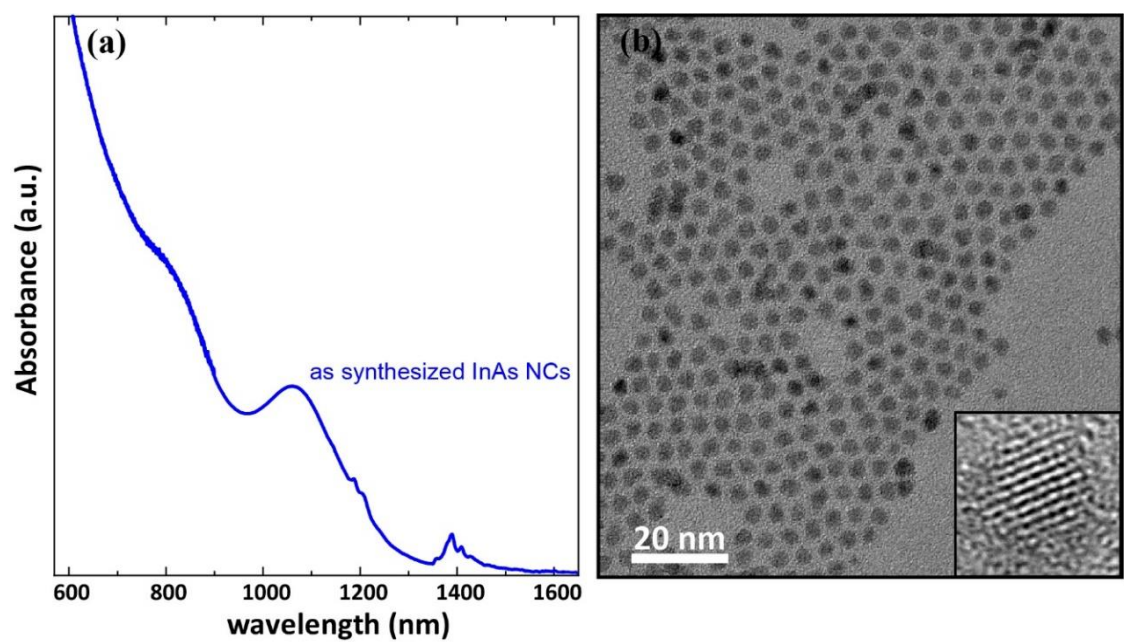

**Figure S1**: High resolution SEM image of InAs NCs film annealed at 220c for 30 min. under N₂ environment and its corresponding energy dispersive X-ray spectroscopy spectrum indicating successful ligand exchange due to signal of sulfur.

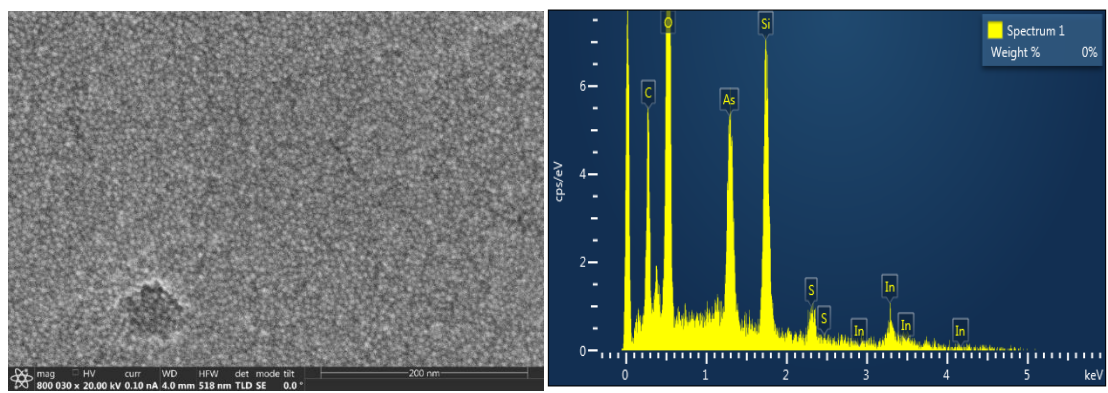

**Figure S2**: High resolution SEM image of InAs NCs film annealed at 220c for 30 min. under N₂ environment and its corresponding energy dispersive X-ray spectroscopy spectrum with signal of sulphur.

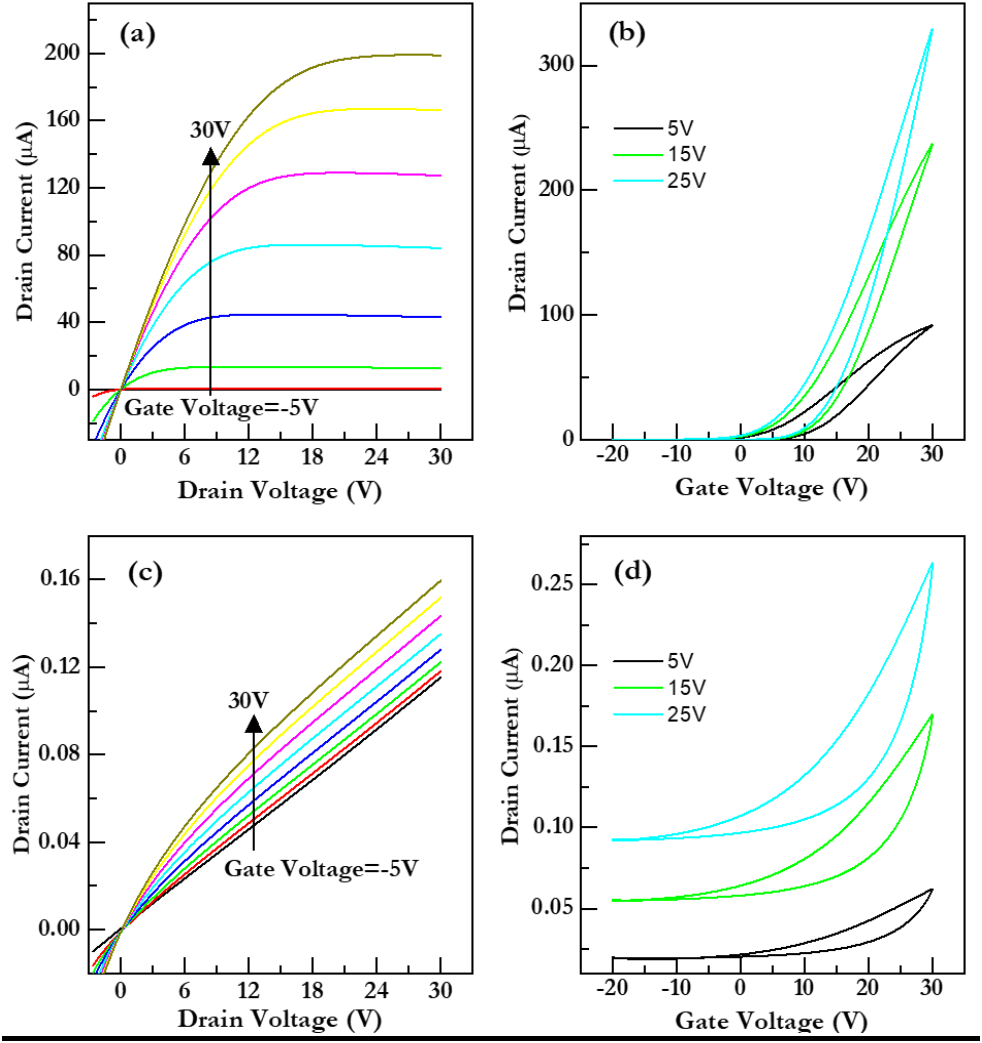

**Figure S3**: Measured output and transfer characteristics of 14.5 Cu/NC doped InAs NC FETs (a, b) and 290 Cu/NC doped InAs NC FETs (c, d) with W/L~200, L~100 µm, $d_{NC}$=35nm and $d_{SiO2}$=100 nm.

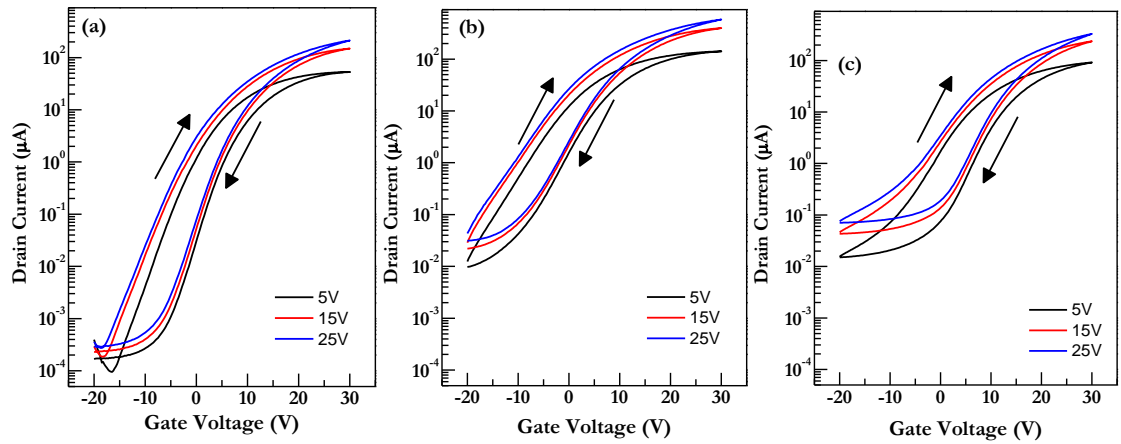

**Figure S4**: Semi-log plots of the transfer characteristics of undoped (a); 1.8 Cu/NC doped (b) and 14.5 Cu/NC doped (c) InAs NC FETs with W/L~200, L~100 µm, $d_{NC}$=35nm and $d_{SiO2}$=100 nm.

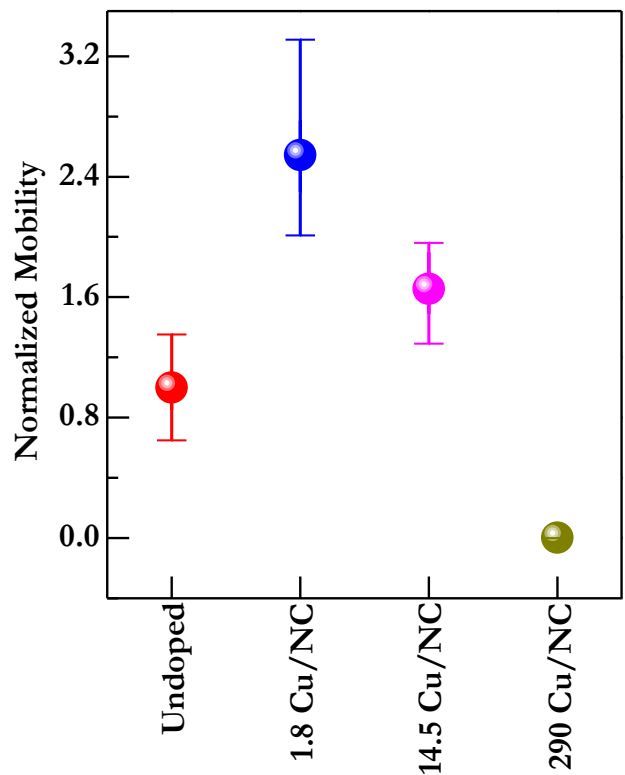

**Figure S5**: Statistics of mobility variation as a function of doping concentration for six transistors on different wafers fabricated in two different runs. The mobility values are normalized with the average mobility of undoped InAs transistors.

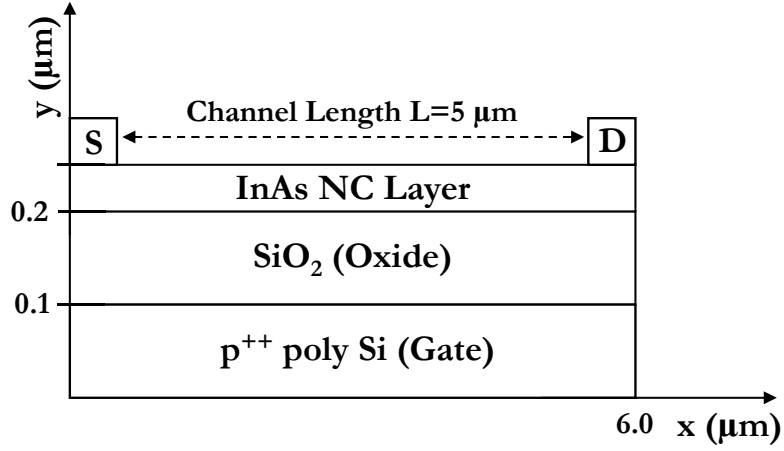

**Figure S6:** Schematic diagram of the top-contact NC structure for which two-dimensional simulations of the electron transport are performed using the Sentarus TCAD platform.

**Table S1:** List of NC layer parameters used in all calculations in this paper.

| Parameters | Value |
|---|---|
| layer Thickness | 35 nm |
| Electron Affinity | 4.05 eV |
| Bandgap | 1.1 eV |
| Dielectric constant | 10 |
| $N_c$ | $2.54 \times 10^{19}$ cm$^{-3}$ |
| $N_v$ | $2.54 \times 10^{19}$ cm$^{-3}$ |
| Work Function S/D | 4.3 eV |